\documentclass[aps,pra,amsmath,amssymb,amsfonts,superscriptaddress,floatfix,nofootinbib]{revtex4} 
\pdfoutput=1
\usepackage{amsmath,amsthm,amsfonts,amssymb}
\usepackage{graphicx}

\usepackage{color}
\usepackage{hyperref}
\newtheorem{conjecture}{Conjecture}

\newtheorem{lemma}{Lemma}

\usepackage{algorithm}
\usepackage[capitalize,nameinlink]{cleveref}

\newcommand{\be}{\begin{equation}}
\newcommand{\ee}{\end{equation}}

\newcommand{\cO}{{\cal O}}
\newcommand{\cS}{{\cal S}}

\newcommand{\expec}{\mathbb{E}}

\newcommand{\hc}{{\rm h.c.}}
\newcommand{\geqsos}{\underset{\rm SoS}{\geq}}
\begin{document}

\title{Perturbation Theory and the Sum of Squares}

\author{Matthew B.~Hastings}

\affiliation{Station Q, Microsoft Research, Santa Barbara, CA 93106-6105, USA}
\affiliation{Microsoft Quantum and Microsoft Research, Redmond, WA 98052, USA}
\begin{abstract}
The sum-of-squares (SoS) hierarchy is a powerful technique based on semi-definite programming that can be used for both classical and quantum optimization problems. 
This hierarchy goes under several names; in particular, in quantum chemistry it is called the reduced density matrix (RDM) method.
We consider the ability of this hierarchy to reproduce weak coupling perturbation theory for three different kinds of systems: spin (or qubit) systems, bosonic systems (the anharmonic oscillator), and fermionic systems with quartic interactions.
For such fermionic systems, we show that degree-$4$ SoS (called $2$-RDM in quantum chemistry) \emph{does not} reproduce second order perturbation theory but degree-$6$ SoS ($3$-RDM) \emph{does} (and we conjecture that it reproduces third order perturbation theory).  Indeed, we identify a fragment of degree-$6$ SoS which can do this, which may be useful for practical quantum chemical calculations as it may be possible to implement this fragment with less cost than the full degree-$6$ SoS.  Remarkably, this fragment is very similar to one studied in \cite{hastings2021optimizing} for the Sachdev-Ye-Kitaev (SYK) model.
\end{abstract}
\maketitle

The sum-of-squares (SoS) hierarchy is a powerful technique for studying both classical and quantum Hamiltonians.  
This hierarchy goes by a variety of names.  
For classical optimization problems, the famous Goemans-Williamson algorithm\cite{GW95,CW04} for MAX-CUT is an example of the lowest level of the hierarchy.
In quantum chemistry, a similar hierarchy for fermions is known as the reduced density matrix method (RDM)\cite{coleman1963structure,erdahl1978representability,
percus1978role,mazziotti2001uncertainty,Nakata_2001,Maz12,klyachko2006quantum}.  There is also a non-commutative SoS hierarchy for qudit systems\cite{HM04,NPA08,DLTW08,PNA10}.

In this paper, we focus on the quantum case.
We consider several cases: qubits (``spins"), fermions, and continuous variables.  We use a parameter $N$ to describe the number of qubits or fermionic modes.
These methods are useful because they can be implemented in time polynomial in $N$ at any given level of the hierarchy, and they can give good approximate answers.  

However, while one interest in these methods is their power to give \emph{non-perturbative} information, a simple check on the methods is to ask: can they reproduce perturbation theory at weak coupling?  That is, if the Hamiltonian of interest is the sum of some simple terms (a quadratic term for fermions, for example) plus some weak perturbation, do they correctly describe the behavior to some given order of perturbation theory?

We emphasize that this question about perturbation theory is particular to quantum systems rather than classical systems.  For example, consider a quantum Hamiltonian $H=-\sum_i Z_i+U$, where we have a system of qubits with $Z_i$ being the Pauli $Z$ operator on the given qubits, and $U$ being some weak perturbation which is a low order polynomial in the Pauli operators.  Then, if $U=0$, the ground state has all spins polarized in the $Z$ direction.  If $U$ is nonzero, but $U$ is diagonal in the $Z$ basis (i.e., $U$ describes a classical perturbation), then the ground state is unchanged for small enough $U$, i.e., such a weak classical perturbation has no effect.  However, a weak perturbation which is off-diagonal in the $Z$-basis, such as an $X$ magnetic field, will change the ground state.

In this paper, we consider this question, and we
 further identify ``fragments" of the hierarchy which are capable of reproducing certain features of perturbation theory, thus suggesting that those fragments may be worth practical application and further exploration, especially in quantum chemistry.

\subsection{Review of Hierarchy}
We very briefly review the hierarchy.  See references above for more.  
In the greatest generality, pick some finite set $\cS$ of operators, and write the operators in this set as $O_a$, where
$a$ is some discrete index.
Then, one considers the matrix $M$ of expectation values of products $O_a^\dagger O_b$, i.e., that is the entry in row $a$ and column $b$.
This matrix $M$ is Hermitian.
In the usual SoS hierarchy, there is some
parameter that we write $d$, called the degree.  This parameter determines the particular set $\cS$ of operators that one consider: the set is taken to be monomials (either in the Pauli operators on individual spins for a spin system or in the creation and annihilation operators for a fermionic system) of degree at most $d/2$,
so that matrix $M$ has size
$\cO(N^{d/2})$-by-$\cO(N^{d/2})$ and it encodes all correlation functions of operators
which are at most degree $d$.

Such a matrix $M$ is subject to various linear relations, determined by commutation and anti-commutation relations of the operators.  For example, given a qubit, the expectation value of $Z^2$ equals $1$, and similarly for $X^2$ and $Y^2$, so the diagonal entries of the matrix are equal to $1$.  The expectation value of $[X,Y]$ is equal to the expectation value of $2iZ$, which gives another linear constraint.
To express such a constraint, we write entries of $M$ as $M_{O_a,O_b}$ to indicate the expectation value of $O_a^\dagger O_b$.  Then, the commutator $[X,Y]=2iZ$ imposes that $M_{X,Y}-M_{Y,X}=2iM_{1,Z}=2iM_{Z,1}$, while the anti-commutator $\{X,Y\}=0$ imposes that $M_{X,Y}+M_{Y,X}=0$.
The full set of constraints for qubits are those imposed by the Pauli commutation and anti-commutation relations.  The full set of constraints for fermionic systems with creation and annihilation operators $\psi^\dagger_i,\psi_i$ are those imposed by the canonical anti-commutation relations: $\{\psi_i,\psi_j\}=\{\psi^\dagger_i,\psi^\dagger_j\}=0$ and $\{\psi^\dagger_i,\psi_j\}=\delta_{i,j}$.
The full set of constraints for continuous variables $p_j,q_j$ which are canonically conjugate are those imposed by the Heisenberg commutation relations, $[q_j,p_k]=i\delta_{j,k}$.

Further, in addition to these linear constraints, any such matrix $M$ must be positive semi-definite.
The SoS hierarchy at degree $d$ corresponds to a relaxation in which these linear constraints and the positive semi-definite constraint are the \emph{only} constraints imposed on the matrix.  
We will call any matrix obeying these constraints a ``pseudo-expectation", and we write $\expec[O]$ to denote the pseudo-expectation value of an operator $O$ of degree at most $d$.
Minimizing the pseudo-expectation value of the Hamiltonian leads to a semi-definite program of polynomial size.  The minimum of this semi-definite program gives a \emph{lower bound} on the true ground state energy of the Hamiltonian.  For sufficiently large $d$, this lower bound becomes equal to the exact ground state energy.

Like any semi-definite program, there is a corresponding dual formulation (the optimization over $M$ is called the ``primal" problem).  In this case, the dual formulation expresses the Hamiltonian as a scalar plus a sum of squares:
one finds operators $O_i$ which are of degree at most $d/2$ such that
\be
\label{HasSoS}
H=\sum_a B_a^\dagger B_a+\lambda_0,
\ee
for some scalar $\lambda_0$.  This then gives a proof that the ground state energy is at least $\lambda_0$, and further that degree-$d$ SoS can prove this as $\expec[B_a^\dagger B_a]\geq 0$ for any pseudo-expectation.
The linear relations for the pseudo-expectation values are used to prove that this expression \cref{HasSoS} of $H$ as a sum of squares holds; for example, as we use later, suppose $H=Z$.  Then, we can express $H=Z=(1/2)(X+iY)(X-iY)-1$, using the commutation relation for $X,Y$.

Remark: whenever we refer to a sum of squares for a quantum system with non-commuting variables, we mean a sum such as $\sum_a B_a^\dagger B_a$.  For commuting variables, of course one can write this separately as a sum of the squares of the Hermitian parts of $B_a$ plus a sum of the squares of $i$ times the anti-Hermitian parts of $B_a$ for for commuting variables it suffices to consider just a sum of squares of Hermitian operators.

As the degree $d$ increase, the cost to solve the semi-definite program increases.  At the same time, sometimes increasing $d$ can lead to \emph{qualitative} improvements in the accuracy of the bound.
As one example, it was found that\cite{hastings2021optimizing} for the so-called SYK (Sachdev-Ye-Kitaev) model\cite{SY93,Kit15} of fermions with degree-$4$ interactions, degree-$4$ SoS was not able to reproduce the correct scaling of the ground state energy with $N$.  However, degree-$6$ SoS does reproduce the correct scaling.  Indeed, a ``fragment" of degree-$6$ reproduces the correct scaling.

\subsection{Results and Outline}
Here we consider a different question: which degree of SoS can correctly reproduce a given order of perturbation theory?  
This means the following.  Suppose we have some parameter $J$ which determines the strength of some perturbation.  We say that a given degree of SoS reproduces the $k$-th order of perturbation theory if the lower bound on the ground state energy computed by that degree of SoS agrees with the exact ground state energy up to terms which are $o(J^k)$, i.e., which are asymptotically smaller than $J^k$.  Indeed, in all cases we study, we will find that when it reproduces the $k$-order, it agrees with the exact ground state energy up to terms which are $O(J^{k+1})$. 

We find that in general, for fermions with quartic interactions, degree-$4$ SoS is \emph{not} able to reproduce the results of second order perturbation theory (i.e., the energy does not have the correct quadratic dependence on the strength of the degree-$4$ terms in the limit that the strength of those terms vanishes).  However, degree-$6$ SoS does reproduce both second order perturbation theory.
Indeed, even a ``fragment" of degree-$6$ SoS suffices.  This is surprisingly similar to \cite{hastings2021optimizing} as it highlights both the weakness of degree-$4$ SoS and the power of degree-$6$ SoS.

In fact, it is the same kind of ``fragment" of degree-$6$ SoS that suffices in both cases.  So, this suggest that this ``fragment" may be useful for practical problems, given that its power has been shown both in weakly interacting systems (i.e., those governed by perturbation theory) and in strongly interacting systems (the SYK model is in some sense regarded as a maximally srongly interacting system).

Let us explain this fragment.  This fragment is obtained by letting the set $\cS$ include all monomials of degree at most $2$ (i.e., the operators $1,\psi_i,\psi^\dagger_i,\psi_i \psi_j,\psi_i \psi^\dagger_j,\psi^\dagger_i \psi^\dagger_j$) as well as $\cO(N)$ additional operators which are \emph{polynomials} of degree $3$ in $\psi,\psi^\dagger$, i.e., linear combinations of monomials.
We write these additional operators $\tau_i$ and $\tau_i^\dagger$, where the index $i$ ranges over the $\cO(N)$ values, and we include both the operator and its Hermitian conjugate in $\cS$.
We call this a ``fragment" because it includes some, but not all, of the information in the degree-$6$ hierarchy.

The only difference between the fragment that we use in this paper, and that used in \cite{hastings2021optimizing} for the SYK model is a different choice of the particular $\tau_i$.

We include the following linear relations as constraints on the pseudo-expecation:
\begin{itemize}
\item[{\bf 1.}] The submatrix where both row and column correspond to a monomial of degree at most $2$ in $\psi,\psi^\dagger$ is subject to all linear relations that are present in the usual degree-$4$ hierarchy.

\item[{\bf 2.}] Any product of degree $1$ in $\psi$ or $\psi^\dagger$ and degree $1$ in $\tau$ or $\tau^\dagger$, such as $\psi_i \tau_j$, $\psi_i \tau_j^\dagger$, $\psi_i^\dagger \tau_j$, $\psi_i^\dagger \tau_j^\dagger$, is equal to some polynomial of degree $4$ in $\psi,\psi^\dagger$, and hence we impose the corresponding constraint on the matrix.
Indeed, it suffices to implement these relations for the products $\psi^\dagger_i \tau_i$ and $\psi_i \tau_i^\dagger$.

\item[{\bf 3.}] Anti-commutators such as $\{\tau_i,\tau_j\}$, 
$\{\tau_i,\tau_j^\dagger\}$, 
$\{\tau_i^\dagger,\tau_j\}$, or
$\{\tau_i^\dagger,\tau_j^\dagger\}$ are equal to polynomials of degree at most $4$ in $\psi,\psi^\dagger$, and hence we impose the corresponding constraint on the matrix.
Remark: this condition is similar to the so-called ``$T_1$ constraint"\cite{erdahl1978representability} in the RDM method of quantum chemistry; however we only consider some subset of degree-$3$ polynomials, rather than all of them, possibly reducing the time overhead, and the constraint {\bf 2} above is an additional constraint which is not part of the usual $T_1$ constraint.
Indeed, it suffices to implement these relations for
the anti-commutators $\{\tau_i^\dagger,\tau_i\}$.
\end{itemize}

Before considering fermionic systems in \cref{sec:fermion}, we first
spin systems in \cref{sec:spin} because they are slightly simpler, showing that degree-$4$ SoS can reproduce third order perturbation theory for quadratic interactions.  We additionally
consider the anharmonic oscillator in \cref{sec:anharmonic}, highlighting some interesting complications that occur due to unbounded continuous variables.

\section{Spin Systems}
\label{sec:spin}
We consider a Hamiltonian
\be
\label{Hdefspin}
H=\sum_j V_j Z_j + \frac{1}{2}\sum_{j,k} J_{jk} X_j X_k,
\ee
where $V_j,J_{jk}$ are scalars and $Z_j,X_j$ are Pauli matrices on qubit $j$.
Here $J$ is a symmetric matrix whose diagonal entries are $0$; the factor of $1/2$ is to avoid double counting.

We show in this section that degree-$2$ SoS is able to reproduce the results of second and third order perturbation theory in $J$.

Remark: \cref{Hdefspin} includes only terms $XX$ as perturbations; more generally one may also include terms $XY$ and $YY$.  We omit these terms to simplify the notation, but we expect that the same result holds.  We comment later on how one should modify the calculation in that case.

Assume all $V_j>0$, without loss of generality.  Then, a standard
perturbation theory calculation gives that the ground state energy is
\begin{eqnarray}
E_0&=&-\sum_j V_j - \sum_{j<k} \frac{J_{jk}^2}{2(V_j+V_k)}
\\ \nonumber
&&+\sum_{j,k,l \, {\rm distinct}}  J_{jk}\frac{1}{2 (V_j+V_k)}   J_{jl}\frac{1}{2(V_k+V_l)} J_{kl}
\\ \nonumber
&& + \cO(J^4).
\end{eqnarray}

.

We shall use the following technique to compute the lower bound given by degree-$2$ SoS.  We consider a \emph{dual semi-definite program}, which computes a lower bound to the optimum of the SoS relaxation.  Indeed, by strong semi-definite programming duality, the lower bound computed by the dual program is equal to the optimum of the primal, and so we may simply refer to ``the" optimum.  However, any feasible solution of the dual program gives some lower bound on the optimum.  We will find a feasible solution of the dual program that gives a lower bound that agrees with perturbation theory up to $\cO(J^4)$ corrections; then, the optimum must also agree with perturbation theory up to $\cO(J^4)$ corrections since the value from the optimum cannot be larger than the true ground state energy and third order perturbation theory correctly computes that energy up to $\cO(J^4)$.

The feasible solution of the dual program is, as mentioned above, a representation of the Hamitonian as a sum of squares plus scalar:
\be
\label{SoSrep}
H=\lambda_0+c_a \sum_a B_a^\dagger B_a,
\ee
for some operators $B_a$ which are \emph{linear} combinations of the Pauli matrices and for some non-negative scalars $c_a$ (indeed, these scalars $c_a$ may be absorbed into $B_a$ but we keep them for clarity later).

For clarity, in what follows we will in some sense ``derive" the desired feasible solution of the dual problem.  That is, rather than simply writing down 
an appropriate feasible solution for the dual program which agrees with perturbation theory up to $\cO(J^4)$ corrections, we will leave give a more general ansatz for the dual program and then solve for the value of terms in this ansatz.

To warm up, suppose first that $J=0$.  Then, we wish to certify that the ground state
energy is at least $-\sum_j V_j$.  One way is to write
$$H=\sum_j V_j \Bigl(\frac{1-Z_j}{2}\Bigr)^2-\sum_j V_j.$$
While this does certify the ground state energy for $J=0$, we use another method that
generalizes to the case $J\neq 0$.
Instead, we can write
$$H=\sum_j \frac{V_j}{2} (X_j+iY_j) (X_j-iY_j)-\sum_j V_j.$$
Here,
we use the standard convention for Pauli matrices that $X_j Y_j = i Z_j$.

Now we generalize to $J\neq 0$.  We write
$$H=\lambda_0+\sum_j \frac{V_j}{2} B_j^\dagger B_j,$$
where
$$B_j=X_j-iY_j+\sum_{k\neq j} A_{jk} X_k,$$
and $\lambda_0$ will be determined.  Here $A$ is a real matrix that need not be symmetric though we will find that the optimum is for $A$ symmetric.

Remark: if we choose $A$ to be complex, we will generate terms $YX$ and we could also add modify the definition to $B_j=X_j-iY_j+\sum_{k\neq j} A_{jk} X_k+\sum_{k \neq j} B_{jk}Y_k $ to produce $YY$ and $XY$ terms.  We restrict to the simplest case.

Then, we have
\be
\label{sumis}
\sum_{j} B_j^\dagger B_j=\sum_j Z_j + \sum_{j \neq k} K_{jk} X_j X_k +
s,
\ee
where
\be
\label{Kis}
K_{jk}=  V_j A_{jk}+\frac{1}{2}\sum_{l \, {\rm distinct \, from}\, j,k} V_l A_{jl} A_{kl}
\ee
and the scalar $s$ is
$$s=\frac{1}{2}\sum_{j \neq k} V_j A_{jk}^2.$$

We will solve for $A$ so that $K_{jk}+K_{kj}=J_{jk}$ and then $H=\lambda_0+\sum_j \frac{V_j}{2} B_j^\dagger B_j,$ with
$$\lambda_0=-s.$$

{\it Second order perturbation theory---}
We will choose an $A$ that is $\cO(J)$.  Hence, the second term on the right-hand side of \cref{Kis} is $\cO(J^2)$.  So, for the moment, let us ignore this term and consider just the first term; more precisely, we will perturbatively solve for $A$ in a series in $J$.  In this case, we want
$$V_j A_{jk}+V_k A_{kj}=J_{jk}+\cO(J^2)$$ to obtain the correct Hamiltonian.  Further, we want to minimize $s$.  Then, the optimum is obtained by a symmetric $A$, with $$A_{jk}=J_{jk}/(V_j+V_k)+\cO(J^2).$$

We then have \begin{eqnarray}s&=&\frac{1}{2}\sum_{j \neq k} V_j \frac{J_{jk}^2}{(V_j+V_k)^2}\\ \nonumber
&=&\sum_{j<k} \frac{J_{jk}^2}{2(V_j+V_k)}+\cO(J^3),
\end{eqnarray}
agreeing with second order perturbation theory.

{\it Third order perturbation theory---}
We now solve for $A$ up to $\cO(J^3)$ corrections.  We choose to keep $A$ symmetric,
and we keep the diagonal entries of $A$ zero and we set
$$
j \neq k \quad \rightarrow \quad
A_{jk}=\frac{J_{jk}}{V_j+V_k}-\frac{1}{V_j+V_k}\sum_{l \, {\rm distinct \, from}\, j,k} V_l \frac{J_{jl}}{V_j+V_l} \frac{J_{kl}}{V_k+V_l}+\cO(J^3),$$
which gives $K_{jk}+K_{kj}=J_{jk}+\cO(J^3)$.
Then, after some algebra, some of which we reproduce here:
\begin{eqnarray}s&=&\frac{1}{2}\sum_{j \neq k} V_j A_{jk}^2\\ \nonumber
&=&\sum_{j<k} \frac{J_{jk}^2}{2(V_j+V_k)}
\\ \nonumber
&&-\sum_{j,k,l\,{\rm distinct}} V_j\frac{J_{jk}}{(V_j+V_k)^2} V_l
\frac{J_{jl}}{V_j+V_l} \frac{J_{kl}}{V_k+V_l}+\cO(J^4)
\\ \nonumber
&=&
\sum_{j<k} \frac{J_{jk}^2}{2(V_j+V_k)}
\\ \nonumber
&&-\frac{1}{2}\sum_{j,k,l\,{\rm distinct}} \frac{J_{jk}}{V_j+V_k} V_l
\frac{J_{jl}}{V_j+V_l} \frac{J_{kl}}{V_k+V_l}+\cO(J^4)
\\ \nonumber
&=&
\sum_{j<k} \frac{J_{jk}^2}{2(V_j+V_k)}
\\ \nonumber
&&-\frac{1}{4}\sum_{j,k,l\,{\rm distinct}} J_{jk}\frac{1}{V_j+V_k} 
J_{jl}\frac{1}{V_k+V_l} J_{kl}
+\cO(J^4).
\end{eqnarray}

This agrees with third order perturbation theory.

{\it Higher orders---}
We have seen that degree-$2$ SoS reproduces second and third order perturbation
theory.  One may ask: for degree-$k$ SoS (with $k$ even), what order of perturbation theory can it reproduce?  We conjecture $(*)$ that it reproduces perturbation theory up to order $k+1$, with an error that is $\cO(J^{k+2})$.  Let us explain our motivation for this conjecture.

The representation of the Hamiltonian as a sum of squares, as in \cref{SoSrep}, will give a tight bound on the ground state energy only if each $B_j$ annihilates the ground state.  More generally, if each $B_j$ annihilates the ground state up to $\cO(J^p)$ error, for some $p$, then the bound on the ground state energy will be accurate to up $\cO(J^{2p})$ error.

For degree-$k$ SoS, each $B_j$ is a polynomial in the Pauli operators of degree $k/2$.  
For $k=2$, this is a linear combination of Pauli operators and one can choose the linear
combination so that it annihilates the ground state up to $\cO(J^2)$ error.
We conjecture $(**)$ that one can find such polynomials $B_j$ so that each $B_j$ annihilates the ground state energy up to $\cO(J^p)$ for $p=(k/2+1)$,
 and hence one can find polynomials giving a
bound on the ground state energy that is tight to error $\cO(J^{2(k/2+1)})=\cO(J^{k+2})$.

So, the conjecture is $(*)$ will follow from conjecture $(**)$.  Conjecture $(**)$ is a natural guess that, at higher orders, one must increase the order of the polynomial to increase the accuracy with which it annihilates the ground state.  We leave the proof of these conjectures for elsewhere.

\section{Fermion Systems}
\label{sec:fermion}
In this section, we consider fermionic systems, first showing that degree-$4$
SoS does not reproduce second order perturbation theory and then finding a fragment of
degree-$6$ SoS that reproduces second order perturbation theory and conjecturally reproduces third order perturbation theory.

\subsection{Failure of degree-$4$ SoS}
We now show that degree-$4$ SoS does not, in general, reproduce second order perturbation theory (much less higher orders!) for fermion systems with a degree-$4$ interaction, even when the Hamiltonian obeys charge conservation.  Our example Hamiltonian using $7$ complex fermionic modes
is as follows.  We label modes
 by either a single label $0$ or by a pair $i,a$ for $i\in \{1,2,3\}$ and $a\in\{1,2\}$.  We let
\begin{align}
\label{Horig}
H=\,&\psi^\dagger_0 \psi_0
+\sum_{a\in\{1,2\}} \Bigl(\psi^\dagger_{1,a} \psi_{1,a}-\psi^\dagger_{2,a}\psi_{2,a}-\psi^\dagger_{3,a}\psi_{3,a}\Bigr)
\\ \nonumber
&
+\epsilon \sum_{a\in\{1,2\}} (\psi^\dagger_0 \psi^\dagger_{1,a} \psi_{2,a} \psi_{3,a}+\hc),
\end{align}
where $+\hc$ means to add the Hermitian conjugate of the previous term,
and where $\epsilon$ controls the perturbation expansion.
Remark: there are even simpler degree-$4$ Hamiltonians for which degree-$4$ SoS does not reproduce second order perturbation theory if we do not impose a requirement of charge conservation.  However, it is worth demonstrating that this problem arises even with charge conservation imposed so we consider this particular Hamiltonian.

Remark: given that the Hamiltonian has charge conservation, there are two additional constraints that may be added.  Let $n=\psi^\dagger_0 \psi_0+\sum_{a\in\{1,2\}} \sum_{i\in\{1,2,3\}} \psi^\dagger_{i,a} \psi_{i,a}$ denote the number operator.  Then, we may impose the constraints $\expec[n^2]=\expec[n]^2$ and $\expec[n]=n_e$, where $n_e$ is an integer equal to the number of electrons in the ground state (here, $n_e=4$ for small $\epsilon$).  This implies the absence of fluctuations in the number operator and implies that the number operator has a given (integer) value and implies the other ``reduction constraints" typically studied in quantum chemistry.  For \cref{fluctuatingnumber}, we ignore this constraint.  In \cref{nonumberfluc}, we show how we may construct an example which includes this constraint but has the same bad behavior.

\subsubsection{Without constraints on number operator}
\label{fluctuatingnumber}
In this subsubsection, we 
do not impose the constraints $\expec[n^2]=\expec[n]^2=n_e^2$.

The Hamiltonian looks more symmetric if we apply particle-hole conjugation to modes $2,a$ and $3,a$ giving, up to an additive scalar that we drop from here on,
\begin{align}
\label{Hph}
H=\,&\psi^\dagger_0 \psi_0
+\sum_{a\in\{1,2\}}\sum_{i\in \{1,2,3\}} \psi^\dagger_{i,a} \psi_{1,a}
\\ \nonumber
&
+\epsilon \sum_a (\psi^\dagger_0 \psi^\dagger_{1,a} \psi^\dagger_{2,a} \psi^\dagger_{3,a}+{\rm h.c.}).
\end{align}
We emphasize that for the rest of this subsection we use the Hamiltonian $H$ from \cref{Hph}.

We can compute the exact ground state energy of $H$.  The ground state subspace is spanned by three states, written $\Psi_0,\Psi_1,\Psi_2$, where
$\Psi_0$ is the state with all modes empty, and
$\Psi_a$ for $a\in\{1,2\}$ is given by
$\Psi_a=\psi^\dagger_0 \psi^\dagger_{1,a} \psi^\dagger_{2,a} \psi^\dagger_{3,a} \Psi_0$.
In this three-dimensional subspace, the Hamiltonian is the matrix
$$
\begin{pmatrix}
0 & \epsilon & \epsilon \\
\epsilon & 4 & 0 \\
\epsilon & 0 & 4,
\end{pmatrix},
$$
so the ground state energy is equal to
$$E_0=-\frac{1}{2}\epsilon^2+\cO(\epsilon^4).$$

We now show the degree-$4$ SoS does not correctly reproduce this second order term.
This particle-hole conjugated Hamiltonian \cref{Hph} has various conserved quantities.
For example, it inherits a conserved quantity from the particle number conservation
of \cref{Horig}.  However, for what follows, we use that the following operator
\be
\label{Qdef}
Q\equiv\psi^\dagger_0 \psi_0 - \frac{1}{3}\sum_{a\in\{1,2\}}\sum_{i\in\{1,2,3\}}
\psi^\dagger_{i,a} \psi_{i,a}
\ee
commutes with the Hamiltonian.

We now give a degree-$4$ pseudo-distribution $\expec[\cdot]$ that gives a more negative second order correction than the exact one.  
We emphasize that we do not claim that this is the optimal degree-$2$ pseudo-distribution; while we optimize over some parameters, there may be even better pseudo-distributions.

Let $u,v$ be real scalars that we choose later.
We will give the pseudo-distribution $\expec[\cdot]$ by giving some pseudo-expectation values which are \emph{normal ordered}, meaning that all annihilation operation are to the right of creation operators.
Any product of operators which is not normal ordered can be reduced to that case using the canonical anti-commutation relations.

Let
$$
\expec[\psi^\dagger_0 \psi_0]=u$$
and also let
$$\expec[\psi^\dagger_{i,a} \psi_{i,a}]=u$$
for all $i,a$.
Let
all other pseudo-expectation values of bilinears vanish, for example $\expec[\psi^\dagger_{i,a} \psi_{j,b}]=0$ if $i\neq j$ or $a\neq b$.
Further, let
$$\expec[\psi^\dagger_0 \psi^\dagger_{i,a} \psi_{i,a} \psi_0]=\expec[\psi^\dagger_{i,a} \psi^\dagger_{j,b} \psi_{j,b} \psi_{i,a}]=u.$$
for all $i,a$ and for all $j,b$ where the pair $j,b$ is distinct from the pair $i,a$, meaning that $i\neq j$ or $a\neq b$ or both.
This is equivalent to
$$
\expec[n_0 n_{i,a}]=\expec[n_{i,a} n_{j,b}]=u,$$
where $n_0$ and $n_{i,a}$ are number operators.

Finally, let
$$
\expec[\psi^\dagger_0 \psi^\dagger_{1,a} \psi^\dagger_{2,a} \psi^\dagger_{3,a}]=v,$$
and
$$
\expec[\psi_{3,a} \psi_{2,a} \psi_{1,a}\psi_0]=v,$$
where the second equality of course is a consequence of the first equality and the assumption that the pseudo-expectation value of the Hermitian conjugate of some operator is the complex conjugate of the pseudo-expectation value of that operator.

All pseudo-expectation values of a normal ordered monomial, other than those given above or those which can be reduced to that above by permuting the operators and using the canonical anti-commutation relations, are zero.

We can describe this pseudo-distribution in words: for $v=0$, it is the same as the distribution where, with probability $u$, all fermionic modes are occupied, and, with probablity $1-u$, all fermionic modes are unoccupied.  That distribution, for $0\leq u \leq 1$, is an actual distribution, not just a pseudo-distribution. 

So, for now on, we assume $0\leq u \leq 1$.
 However,
we must verify positive definiteness in the case that $v\neq 0$.  We must show
$\expec[O^\dagger O]\geq 0$ for any operator $O$ which is of degree at most $2$.  We emphasize that $O$ may contain both creation and annihilation operators, so that $O^\dagger O$ need not be normal ordered.

Since the pseudo-expectation of a monomial of odd degree vanishes, we may separately consider the case that $O$ is a sum of odd degree terms (in this case, it must be degree $1$) and the case that $O$ is sum of even degree terms.  The case that $O$ is homogeneous of degree $1$ is trivial to show, so we only consider the case of even degree.

Given any monomial $M$, we assign it a \emph{charge} $q$, defined by
$[Q,M]=qM$, where $Q$ is given in \cref{Qdef}.
A monomial of degree $2$ may have charge $\pm 4/3,\pm 2/3,0$, while a monomial of degree $0$ (i.e., a scalar) has charge $0$.
Since the pseudo-expectation of any monomial of non-vanishing charge vanishes,
we may assume that $O$ is a sum of monomials of some given charge $q$.

If $q=\pm 4/3$ or $q=0$, then the pseudo-expectation value is independent of $v$, i.e., we may assume $v=0$.  So, positive definiteness must be satisfied.

So, it suffices to consider the case $\pm q=2/3$.  First consider $q=2/3$.
There are are $21$ degree-$2$ monomials with $q=2/3$, up to permutation of the creation and annihilation operators.  Of these, $9$ are of the form
\be
R_{i,j}\equiv \psi_{i,1}\psi_{j,2}
\ee
for $i,j\in\{1,2,3\}$.
The remaining $12$ operators are
\begin{eqnarray}
\nonumber
M_{i,a}&\equiv& \psi^\dagger_0 \psi^\dagger_{i,a},
\\
\nonumber
N_{i,a}&\equiv & \prod_{j\neq i} \psi_{j,a},
\end{eqnarray}
for $i\in \{1,2,3\}$ and $a\in\{1,2\}$ and where the product on the second line is taken in, for example, increasing order of $j$ (the other order differs by an arbitrary sign) so that $N_{1,a}=\psi_{2,a}\psi_{3,a}$.

One may verify that the pseudo-expectation $\expec[O^\dagger R_{i,j}]$ vanishes if $O$ is any
of these monomials other than $R_{i,j}$.

So, we just consider the pseudo-expectations $\expec[M^\dagger_{i,a} M_{j,b}], \,
\expec[M^\dagger_{i,a} N_{j,b}],\,
\expec[N^\dagger_{i,a} M_{j,b}],\,
\expec[N^\dagger_{i,a} N_{j,b}]$.  One may verify that these all vanish unless $i=j$ and $a=b$, and if
$i=j$ and $a=b$ then the pseudo-expectation is independent of $i,a$.  So, we may further simplify by just assuming that $O$ is a linear combination of $M_{i,a}$ and $N_{i,a}$ for any
given choice of $i,a$.

In this case, it reduces to showing positive semi-definiteness of the two-by-two matrix
$$
\begin{pmatrix}
1-u & v \\
v & u \end{pmatrix}.$$
Taking $v$ small and taking $u=v^2+\cO(v^4)$ gives a positive-semi-definite matrix.

If we instead considered $q=-2/3$, the result would be similar, instead assuming $O$ is a linear combination of $M^\dagger_{i,a}$ and $N^\dagger_{i,a}$ and the same $u$ would again give positive semi-definiteness.

Finally, for the given $u$ we have
\be
\expec[H]=4\epsilon v+7u=4\epsilon v + 7v^2+\cO(v^4),
\ee
where the factor of $4$ in front of $\epsilon v$ is a combination of $2$ (for the two choices of $a$) times $2$ (for the pseudo-expectation value of 
$\psi^\dagger_0 \psi^\dagger_{1,a} \psi^\dagger_{2,a} \psi^\dagger_{3,a}$ plus its Hermitian conjugate).
Optimizing over $v$, taking $v=-2\epsilon/7+\cO(\epsilon^3)$, we have
\be
\expec[H]=-\frac{4}{7}\epsilon^2+\cO(\epsilon^4),
\ee
and since $4/7>1/2$, this does not agree with the exact second order result.

\subsubsection{With constraints on number operator}
\label{nonumberfluc}
We now show how we may slightly modify the example to obtain the same bad behavior as in \cref{fluctuatingnumber}
while also imposing the constraints that $\expec[n^2]=\expec[n]^2=n_e^2$.

First, let's show that the pseudo-expectation above does not obey these constraints on number.
After the particle-hole conjugation giving \cref{Hph}, the number operator becomes
$$n=4+\psi^\dagger_0\psi_0+\sum_a\Bigl(\psi^\dagger_{1,a}\psi_{1,a}-\psi^\dagger_{2,a}\psi_{2,a}-\psi^\dagger_{3,a}\psi_{3,a}\Bigr).$$
The pseudo-expectation above has fluctuations in this number operator if $u\neq 0$.  Suppose $v=0$.  Then, with probability $1-u$ we have all modes empty where $n=4$, and with probability $u$ we have all modes occupied where $n=3$.  Further, if $u\neq 0$, we do not have $\expec[n]=4$.

To rectify this, while giving almost the same example as above, 
we add an additional fermionic mode, labelled $4$, so that there are now a total of $8$ modes.
Before particle-hole transformation,
we use the same Hamiltonian as \cref{Horig}, so that no interaction terms involve this mode $4$.
We again apply particle-hole conjugation to modes $2,3$, giving the same Hamiltonian as in \cref{Hph}.
Now the particle-hole transformed number operator is
$$n=4+\psi^\dagger_0\psi_0+\psi^\dagger_4\psi_4+\sum_a\Bigl(\psi^\dagger_{1,a}\psi_{1,a}-\psi^\dagger_{2,a}\psi_{2,a}-\psi^\dagger_{3,a}\psi_{3,a}\Bigr).$$

We use essentially the same construction of a pseudo-expectation as above, except
we also take
$$\expec[\psi^\dagger_4 \psi_4]=u$$
and
$$\expec[\psi^\dagger_0 \psi^\dagger_4 \psi_4 \psi_0]=\expec[\psi^\dagger_4 \psi^\dagger_{i,a} \psi_{i,a} \psi_4]=u.$$
That is, for $v=0$ it is again the same as the pseudo-expectation where with probability $u$ we have all modes occupied and with probability $1-u$ we have all modes unoccupied.

Now we have $\expec[n^2]=\expec[n]^2=n_e^2$ for $n_e=4$, and this is a valid pseudo-expectation for the same $u,v$ as above.

Further, we have checked this by solving the relevant SDP for the $2$-RDM numerically for this Hamiltonian\footnote{Using code by D. Wecker to solve the SDP.}.  We have compared the SDP solution for the $2$-RDM to the solution of the much simpler SDP given by the ansatz above: we impose that
$$
\begin{pmatrix}
1-u & v \\
v & u \end{pmatrix}$$
is positive semi-definite and
minimize
$4\epsilon v+7u$.  Numerically, the two solutions agree very closely for a range of $\epsilon$.

\subsection{Fragment of degree-$6$ SoS Succeeds}
We now show that degree-$6$ SoS does reproduce second order perturbation theory (we expect that it can also reproduce third order perturbation theory but we do not give the combinatorics).  
We work in the most general case, allow interactions that do not conserve particle number.  While often such interactions are written using a representation in terms of Majorana operators, we will use fermion creation and annihilation operators rather than Majorana operators.

Then, the most general case is
\be
\label{Hgen}
H=\sum_j E_j \psi^\dagger_j \psi_j+\epsilon \sum_{p=0}^4 H_{p,4-p},
\ee
where all $E_j$ are positive scalars, where $\epsilon$ is a small parameter controlling the perturbation theory, and where the other terms in the Hamiltonian are discussed below.
Of course, in a typical problem from physics or chemistry, one will instead often have some $E_j$ positive and some negative.  In that case, one can apply particle-hole conjugation to the orbitals $j$ for which $E_j$ is negative (interchanging $\psi^\dagger_j$ and $\psi_j$) to make the $E_j$ positive.

We define each term $H_{p,4-p}$ to be a sum of products of $p$ creation operators and $4-p$ annihilation operators, with the term normal ordered so that the annihilation operators are to the right of the creation operators.  Thus, all terms $H_{p,4-p}$, except for $H_{4,0}$, annihilate the unperturbed ground state (i.e., when all number operators $n_i$ are equal to $0$).
Remark: terms $H_{p,4-p}$ with $p\neq 2$ can arise even for physical systems with particle-number conserving Hamiltonians, due to the particle-hole conjugation discussed above.

Remark: let us slightly clarify the particle-hole conjugation.  Suppose we consider some Hamiltonian which has not been particle-hole conjugated, so that some $E_j$ are negative and some are positive.  Consider adding some perturbation $\psi^\dagger_j \psi_k+\hc$ which conserves particle number to this Hamiltonian.  If $E_j,E_k>0$, then after particle-hole conjugation this perturbation is unchanged.  This term annihilates the ground unperturbed ground state so perturbatively it has no effect on the ground state energy; indeed, this is not surprising as it corresponds to a hopping between two empty orbitals.  Similarly, if $E_j,E_k<0$, then particle-hole conjugation leaves this term unchanged up to an overall sign, and again the term annihilates the unperturbed ground state.  However, if $E_j>0$ and $E_k<0$, then particle-hole conjugation turns this perturbation into $\psi^\dagger_j \psi^\dagger_k+\hc$ which does not annihilate the unperturbed ground state.

We have
\be
H_{p,4-p}=H_{4-p,p}^\dagger.
\ee
To fix notation, we define
\be
H_{4,0}= \frac{1}{4!}\sum_{j,k,l,m}
V_{j,k,l,m} \psi^\dagger_j \psi^\dagger_k \psi^\dagger_l \psi^\dagger_m,
\ee
for some totally anti-symetric tensor $V$.
Then
\be
H_{0,4}=H_{4,0}^\dagger=
\frac{1}{4!}\sum_{j,k,l,m}
V_{j,k,l,m} \psi_j \psi_k \psi_l \psi_m.
\ee

Remark: given an arbitrary Hamiltonian term which is degree-$4$ in the creation and annihilation operators, we can normal order that term to cast it into the form above.  Doing this normal ordering can generate additional scalars and degree-$2$ terms.  Consider, for example, $\psi_1 \psi_1^\dagger \psi_2 \psi_2^\dagger=(1-n_1)(1-n_2)$, where $n_j$ is the number operator $\psi^\dagger_j \psi_j$.  This equals
$1-n_1-n_2+n_1n_2=1-\psi^\dagger_1 \psi_1-\psi_2^\dagger \psi_2+
\psi^\dagger_2 \psi^\dagger_1 \psi_1 \psi_2$.
So, really when we consider the most general perturbation, we should also include degree-$2$ terms in the Hamiltonian (any scalar term can be dealt with trivially of course).  We do not include these degree-$2$ terms in the perturbation however because {\bf (1)}: they can be dealt with using the same techniques as for the degree-$4$ terms, and including them would clutter the notation {\bf (2)}: these degree-$2$ terms can also be incorporated into a shift in the unperturbed Hamiltonian.  Indeed, incorporating these degree-$2$ terms into the unperturbed Hamiltonian may change the orbital basis which diagonalizes that Hamiltonian, but we may choose a Hartree-Fock basis, and in such a basis these degree-$2$ terms vanish.

Standard perturbation theory results give that the ground state energy is equal to $E^{(2)}+\cO(\epsilon^3)$, where $E^{(2)}$ is the second
order perturbation term given by
\be
E^{(2)}=-\frac{1}{4!}\epsilon^2 \sum_{j,k,l,m} \frac{|V_{j,k,l,m}|^2}{E_j+E_k+E_l+E_m}.
\ee
Note that the ground state energy vanishes at $\epsilon=0$ and note that the term linear in $\epsilon$ also vanishes.
Our goal is to show that degree-$6$ sum of squares can reproduce second order perturbation theory.

We first show the following:
\begin{lemma}
\label{lemadd}
The sum $H_{1,3}+H_{2,2}+H_{3,1}$
can be written as a sum of squares of terms of degree at most $2$, plus some term which is linear in number operators, i.e., some $\sum_i a_i n_i$, where $a_i$ are scalars.

More generally, any Hermitian normal ordered polynomial of even degree $d$ in creation and annihilation operators which annihilates the unperturbed ground state can be written as a sum of squares
of terms at most degree $d/2$ plus
some term which is linear in number operators, i.e., some $\sum_i a_i n_i$, where $a_i$ are scalars.
\begin{proof}
We prove the general case.
Any such polynomial is a sum of normal ordered monomials.  Consider a given such monomial, which we denote $T$.  Suppose $T$ is written in normal ordered form as
$c \phi_1 \phi_2 \ldots \phi_d$, where each $\phi_a$ is a creation or annihilation operator
and $c$ is a complex scalar.
Define
$$\Psi_1^\dagger\equiv \phi_1 \ldots \phi_{d/2},$$
and
$$\Psi_2 \equiv \phi_{d/2+1} \ldots \phi_d.$$
Then,
$$T+\hc=c\Psi_1^\dagger \Psi_2+\hc.$$
  Note
$\Psi_1,\Psi_2$ are normal ordered monomials 
of degree $d/2$, and $\Psi_1,\Psi_2$ both annihilate the unperturbed ground state.
We have
\begin{eqnarray}
T&=&(\Psi_1^\dagger+\overline c\Psi_2^\dagger)(\Psi_1+c\Psi_2)
-\Psi_1^\dagger \Psi_1 -|c|^2 \Psi_2^\dagger \Psi_2.
\end{eqnarray}
Consider a term such as 
$-\Psi_1^\dagger \Psi_1$ or $-|c|^2 \Psi_2^\dagger \Psi_2$.
Both these terms are equal to a negative scalar coefficient (either $-1$ or $-|c|^2)$ times some product of $d/2$ operators, with each of those $d/2$ operators being either some number operator $n_j$ for some $j$ or being $1-n_j$ for some $j$.  

In the particular case of most interest, where $d=4$, it is easy to show that
such a term is a sum of squares of polynomials in the creation and annihilation operators of degree at most $2$ plus some linear combination of number operators; for example, $-n_1 n_2=(n_1-n_2)^2/2-n_1/2-n_2/2$.
However, for completeness, we give a general proof.

First, any such term is a real polynomial in the number operators of degree at most $d/2$ which vanishes in the unperturbed ground state (i.e., the term of degree $0$ in the number operators vanishes).
The proof that such a polynomial is a sum of squares plus linear combination of number operators is inductive: we show that any monomial in the number operators of any given degree $d'>1$ (note that here $d'$ refers to the degree in the number operators, so it is degree $2d'$ in the creation and annihilation operators) can be written as a sum of squares of polynomials in the creation and annihilation operators, where each term being squared has degree at most $d'$ in the creation and annihilation operators, plus some polynomial in the number operators of degree at most $\lceil d'/2 \rceil<d'$ such that that polynomial vanishes on the unperturbed ground state.

To show the induction step, consider any such monomial, $c n_{j_1} n_{j_2} \ldots n_{j_{d'}}$ for some coefficient $c$.  Let $T$ denote the given monomial.  If $c>0$ then the result is immediate: 
$T=(\sqrt{c} \psi_{j_1} \ldots \psi_{j_{d'}})^\dagger(\sqrt{c} \psi_{j_1} \ldots \psi_{j_{d'}})$.
Suppose instead $c<0$.  Then we consider two cases.  In the first case, $d'$ is even.
Then $$T=B^\dagger B+\ldots,$$
where
$$B=B^\dagger=\prod_{a=1}^{d'/2} n_{j_a}+\frac{c}{2}\prod_{a=d'/2+1}^{d'} n_{j_a},$$
and
where
$\ldots$ denotes some polynomial in the number operators of degree $d'/2$ which vanishes in the unperturbed ground state.  In this case, the polynomial is $-\prod_{a=1}^{d'/2} n_{j_a}-\frac{c^2}{4}\prod_{a=d'/2+1}^{d'} n_{j_a}.$
In the second case, $d'$ is odd.
Then
$$T=F^\dagger F+\ldots,$$
where
$$F=\psi_{d'} \Bigl(\prod_{a=1}^{(d'-1)/2} n_{j_a}+\frac{c}{2}\prod_{a=(d'-1)/2+1}^{d'-1} n_{j_a}\Bigr),$$
and where again
$\ldots$ denotes some polynomial in the number operators of degree $\lceil d'/2\rceil$ which vanishes in the unperturbed ground state.
Remark: of course in the first case one may instead consider $B=B^\dagger=\lambda \prod_{a=1}^{d'/2} n_{j_a}+\frac{c}{2\lambda}\prod_{a=d'/2+1}^{d'} n_{j_a}$ for any scalar $\lambda\neq 0$, and similarly in the second case.
\end{proof}
\end{lemma}

The reason for this lemma is that it means that it suffices to show that degree-$6$
sum of squares can reproduce second order perturbation theory for a Hamiltonian
\be
\label{Hsimp}
H=\sum_j E_j \psi^\dagger_j \psi_j
+\epsilon \Bigl( H_{0,4}+H_{4,0}+\sum_j a_j \psi^\dagger_j \psi_j\Bigr),
\ee
for some scalars $a_j$, as any representation of \cref{Hsimp} as a sum of squares plus scalar gives a representation of \cref{Hgen} as a sum of squares plus scalar (simply add the sum of squares given by \cref{lemadd}), and both Hamiltonians have the
same second order perturbation theory.

In what follows, we ignore the term $\epsilon\sum_j a_j \psi^\dagger_j \psi_j$ in \cref{Hsimp}, i.e., we set $a_j$ to $0$.  
Indeed, if $a_j \neq 0$, we can incorporate $a_j$ into a shift in $E_j$ by an amount $\cO(\epsilon)$.  One can then incorporate such a shift into our analysis below (i.e., replace all occurrences of $E_j$ by $E_j+\epsilon a_j$), and perturbatively expand in $\epsilon$, seeing that this will only change the result at order $\cO(\epsilon^3)$ and higher.

The reader may wonder why we do not use a similar trick to \cref{lemadd} for terms in $H_{4,0}+H_{0,4}$ which remain in \cref{Hsimp}.  Consider some term $T=\psi^\dagger_i \psi^\dagger_j \psi^\dagger_k \psi^\dagger_l + \psi_i \psi_j \psi_k \psi_l$.  We can set
$A=\psi^\dagger_j \psi^\dagger_i$ and $B=\psi_k \psi_l$ and then
$T=(A^\dagger +B^\dagger)(A+B)-A^\dagger A -B^\dagger B$.  However,
$A^\dagger A=(1-n_i) (1-n_j)$ and does not have vanishing expectation value in the unperturbed ground state.  So, while this would give some correct lower bound on the energy, the bound would give a \emph{linear} change in the ground state energy with $\epsilon$ and so would not reproduce second order perturbation theory.

So, for the rest of the section we consider a simplified Hamiltonian, that we call $H_0$, given by
\be
\label{Hsimp2}
H_0=
\sum_j E_j \psi^\dagger_j \psi_j+
\epsilon \Bigl(\frac{1}{4!}\sum_{j,k,l,m}
V_{j,k,l,m} \psi^\dagger_j \psi^\dagger_k \psi^\dagger_l \psi^\dagger_m+\hc\Bigr).
\ee
Let
\be
\tau_j\equiv \frac{1}{4!}\sum_{k,l,m} \frac{V_{j,k,l,m}}{E_j+E_k+E_l+E_m} \psi^\dagger_k \psi^\dagger_l \psi^\dagger_m.
\ee
Let
\be
F_j=\psi_j+\epsilon\tau_j.
\ee
Then,
\be
\sum_j E_j F_j^\dagger F_j
=H+\epsilon^2 \sum_j E_j  \tau_j^\dagger \tau_j.
\ee
So,
\be
\sum_j E_j F_j^\dagger F_j + \epsilon^2 \sum_j E_j \tau_j \tau_j^\dagger
=H+\epsilon^2 \sum_j E_j  \{\tau_j^\dagger,\tau_j\}.
\ee
We now consider the anti-commutators in
$$\{\tau_j^\dagger,\tau_j\}=\Bigl(\frac{1}{4!}\Bigr)^2\sum_{k,l,m} \sum_{n,o,p}
\frac{V_{j,k,l,m}}{E_j+E_k+E_l+E_m}
\frac{\overline V_{j,n,o,p}}{E_j+E_n+E_o+E_p} 
\{
 \psi^\dagger_k \psi^\dagger_l \psi^\dagger_m,
 \psi_n \psi_o \psi_p\}.$$
For given $k,l,m,n,o,p$ we consider the anti-commutator
$\{
 \psi^\dagger_k \psi^\dagger_l \psi^\dagger_m,
 \psi_n \psi_o \psi_p\}.$
 We assume $k,l,m$ are distinct and $n,o,p$ are distinct.
 There are four cases, depending on the cardinality of the intersection of the set
 $\{k,l,m\}$ with the set $\{n,o,p\}$.  In the first case, this cardinality is $0$, i.e., $k,l,m,n,o,p$ are all distinct.  In this case, the anti-commutator vanishes.
 In the second case, the cardinality is $1$.  Without loss of generality, we assume that
 $m=n$ and $k,l,m,o,p$ are distinct.
 In this case, the anti-commutator is $\psi^\dagger_k \psi^\dagger_l \psi_o\psi_p$.
 In the third case, the cardinality is $2$.  Without loss of generality, we assume that
 $l=o, m=n$ and $k,l,m,p$ are distinct.  Then, the anti-commutator is
 $\psi^\dagger_k \psi_p (n_n+n_o-1)$.
 In the fourth case, the cardinality is $3$.  Without loss of generality we assume that $k=p,l=o,m=n$.  Then, the anti-commutator is
 $n_n n_o n_p + (1-n_n) (1-n_o) (1-n_p)$.
 
 We may summarize these cases as follows: in all except the fourth case, the anti-commutator is a normal ordered polynomial in the fermi operators, containing terms of degree $2$ and $4$ which annihilate the unperturbed ground state (indeed, each term has the same number of creation operators as annihilation operators in it).
 Further, 
in the fourth case, the anti-commutator is equal to $1$ plus some normal ordered polynomial of the fermi
operators, again containing terms of degree $2$ and $4$ which annihilate the unperturbed ground state.
So if we define
\begin{eqnarray}
W&=&\epsilon^2 \sum_j E_j  \{\tau_j^\dagger,\tau_j\}-\frac{1}{4!}\sum_{j,k,l,m} \frac{|V_{j,k,l,m}|^2}{E_j+E_k+E_l+E_m}
\\ \nonumber
&=&
\epsilon^2 \sum_j E_j  \{\tau_j^\dagger,\tau_j\}+E^{(2)},
\end{eqnarray}
then $W$ can be written as a normal order polynomial which annihilates the ground state, i.e., we have cancelled the scalar term by adding $E^{(2)}$.

Then
\be
\sum_j E_j F_j^\dagger F_j + \epsilon^2 \sum_j E_j \tau_j \tau_j^\dagger-W
=H-E^{(2)}.
\ee
Since $-W$ is a normal ordered polynomial in creation and annihilation operators which annihilates the ground state, by \cref{lemadd} the polynomial $-W$ is a sum of squares of degree at most $2$, plus some linear combination of number operators of order $\epsilon^2$.
Thus
$H-E^{(2)}$ is a sum of squares plus some term linear in the number operators; as above, we can incorporate that linear term into a shift in the $E_j$.

This completes the proof that the given fragment of degree-$6$ SoS can reproduce second order perturbation theory.

\section{Continuous Variables}
\label{sec:anharmonic}
Of course, having considered spin systems and fermionic systems, a natural case to consider is bosonic systems with continuous variables.  For example, the anharmonic oscillator
\be
H=\frac{p^2+q^2}{2}+g q^4,
\ee
where $q,p$ are position and momentum variables subject to the commutation relation $[q,p]=i$.  More generally, we may several continuous variables, subject to $[q_j,p_k]=i\delta_{j,k}$.

Immediately we must consider the following point: even if we consider polynomials in several \emph{commuting} real variables, there are polynomials which are everywhere positive, but which cannot be expressed as a sum of squares of polynomials.
This contrasts with the case of spin systems (including the case of classical Boolean variables) and the case of fermionic systems where every non-negative polynomial can be written as a sum of squares\footnote{Proof: if $H\geq 0$, then $H=(\sqrt{H})^2$ and $\sqrt{H}$ is Hermitian.  Any operator, including $\sqrt{H}$, can be written as a polynomial for a spin or fermionic system.}.

However, in the case of commuting real variables, every polynomial which is everywhere non-negative can be written as a sum of squares of ratios of polynomials, i.e., if polynomial $P$ is everywhere non-negative, then $P=\sum_a (r_a/s_a)^2$ for some polynomials $r_a,s_a$.  Equivalently, there is some polynomial $Q$ such that $Q$ is nowhere vanishing and such that $PQ^2$ is a sum of squares of polynomials: one may choose $Q$ to be the product of all the $s_a$.
This is the famous positivstellensatz.

\subsection{Conjectures}
We make the following conjecture for polynomials in multiple position and momentum variables as a generalization. 
\begin{conjecture}
\label{conjrsos}
Given any Hermitian operator $H$ which is polynomial in variables $p_j,q_k$ with $[p_j,q_k]=i\delta_{j,k}$, such that $H$ is positive definite,
then there exists 
an invertible operator $L$ (not necessarily Hermitian),
such that
$L^\dagger H L$ is a sum of squares of polynomials, i.e., $L^\dagger H L = \sum_a B_a^\dagger B_a$, for some $B_a$ polynomial in variables $p_j,q_k$.
\end{conjecture}
Remark: since $L$ is invertible, the existence of such an $L$ gives a certificate that $H$ is non-negative: $H=(L^{-1})^\dagger \sum_a (B_a^\dagger B_a) L^{-1}$.

Remark: a weaker conjecture is that
$$H=\sum_a (L_a^{-1})^\dagger N_a^\dagger N_a L_a^{-1},$$ for some $L_a,N_a$
which are polynomial in $p_j,q_k$ and such that all $L_a$ are invertible.
In the commuting case, this weaker conjecture is equivalent to the above conjecture: one may take $L=\prod_a L_a$ and take $B_a=N_a L/L_a$.  However, with non-commuting variables, the conjectures are not obviously equivalent.

Remark: the reader may note that we have assumed in the conjecture that $M$ is positive definite rather than merely positive semidefinite which is what is assumed in the commuting case.  This is based on the following consideration:
suppose that the result held for positive definite $M$.  Then, for any given $g$, there would be some semidefinite program whose optimum gave the ground state energy.  However, if $g$ is a rational number, then this is a semidefinite program whose coefficients are rational, and hence whose optimum is an algebraic number\cite{Nie_2008}.  However, it seems unlikely that the ground state energy is algebraic for all rational $g$.

\subsection{Anharmonic Oscillator}
We now consider the application of SoS to the anharmonic oscillator
\be
H=\frac{p^2+q^2}{2}+g q^4.
\ee
The conjecture(s) above suggest that we should seek some $L$ and represent $L^\dagger H L$ as a sum of squares.  However, let us first try to represent $H$ itself as a sum of squares.  We will see that this does not give good answers; indeed, it cannot (at any order!) even reproduce first order perturbation theory.

Suppose
$H=\sum_a B_a^\dagger B_a+\lambda_0,$ for some $\lambda_0$ and some polynomials $B_a$.
We immediately see that every $B_a$ must be a polynomial of degree at most $2$, as $H$ is of degree $4$.
Indeed, further any monomial of degree $2$ can only be equal to $q^2$, not $pq$ or $p^2$.

Let us say that an \emph{even} polynomial is in the span of monomials of even degree, i.e., considering just those of degree at most $2$, these are in the span of $1,p^2,pq,q^2$, and considering just those that may occur in $B_a$ those are in the span of $1,q^2$.  Let us say that an \emph{odd} polynomial is in the span of monomials of odd degree, i.e., considering just those of degree at most $2$, these are in the span of $p,q$.

Note that $H$ has even degree.  We claim that we can assume that each $B_a$ is either even or odd.
Indeed, let each $B_a$ be a sum of both even and odd terms: $B_a=E_a+O_a$ for even $E_a$ and odd $O_a$.
Then, $\sum_a E_a^\dagger E_a + \sum_a O_a^\dagger O_a$ agrees with $\sum_a B_a^\dagger B_a$ up to terms of odd degree, and since these terms vanish in $H$ they must cancel in $\sum_a B_a^\dagger B_a$, so indeed
$H=\sum_a E_a^\dagger E_a + \sum_a O_a^\dagger O_a$

Consider the sum
$\sum_a E_a^\dagger E_a$.  This is a polynomial 
$x+yq^2+zq^4$ for some real scalars $x,y,z$.  We have $z=g$.  We wish to take $x$ as small as possible to obtain the optimum bound so we may assume the polynomial equals $0$ for some $q^2$.  We will see that we want $y<0$ so we may assume that the minimum occurs for $q^2>0$. So, since the polynomial is a quadratic in $q^2$, we may assume that indeed the polynomial is a square $\sum_a E_a^\dagger E_a=(C+Dq^2)^2$.

So,
we want to represent
\be
\label{anharmrep}
H=\sum_a O_a^\dagger O_a+(C+Dq^2)(C+Dq^2)+\lambda_0.
\ee

Immediately we have
$D=\pm\sqrt{g}$, and we will see that the optimum is
with $CD<0$.  So, let us take $C>0$ and $D=-\sqrt{g}$.

Then, $H-(C+Dq^2)(C+Dq^2)$ is of harmonic oscillator form: it equals
$p^2/2+(1+4C\sqrt{g})q^2/2-C^2$.
So, $H-(C+Dq^2)(C+Dq^2)$ has ground state energy $(1/2) \sqrt{1+4C\sqrt{g}}$, and this may be shown using a single odd term in the sum of squares: 
$H-(C+Dq^2)(C+Dq^2)=(1/2) \sqrt{1+4C\sqrt{g}}+O^\dagger O$ plus scalar, for
$O=Ap+iBq$ for some scalars $A,B$.

So,
$$\lambda_0={\rm max}_C \Bigl( \frac{1}{2}\sqrt{1+4C\sqrt{g}}-C^2\Bigr).$$
For small $\sqrt{g}$ this is
$$\lambda_0={\rm max}_C \Bigl(\frac{1}{2}+C\sqrt{g}-C^2+\cO(C^2 \sqrt{g})\Bigr),$$
giving
$$\lambda_0=\frac{1}{2}+\frac{g}{4}+\cO(g^2).$$

On the other hand, from linear perturbation theory, the true ground state energy is
$1/2+g \langle q^4 \rangle=1/2+3g/4$, where $\langle \ldots \rangle$ denotes a ground state expectation value.
Thus, no order of SoS can reproduce even linear perturbation theory.

At first, the result may seem paradoxical: our optimization of the dual program corresponds to a solution of the primal with $\expec[q^4]=\expec[q^2]^2$.
Consider, however, the matrix of expectation values.  Consider  the $2$-by-$2$ submatrix $\tilde M$ with rows corresponding to the operators $1,q^2$ and similarly for columns (i.e., $\tilde M$ is a submatrix of the matrix $M$ of expectation values).  Then, if $\expec[q^4]=\expec[q^2]^2$,
the submatrix $\tilde M$ must have a zero eigenvalue.  However, the commutator $[q^2,pq]$ is proportional to $q^2$, and so there must be some off-diagonal matrix elements in the entries $M_{q^2,pq}$ or $M_{qp,q^2}$, i.e., with row corresponding to $q^2$ and column corresponding to $pq$, or vice-versa.  These seems then that this would prevent the matrix from being positive semi-definite, as the eigenvector with zero eigenvalue of the given $2$-by-$2$ block has a nonvanishing matrix element to the subspace outside that block.  However, the resolution is simple: we can make $\expec[q^4]$ arbitrarily close to $\expec[q^2]^2$, hence making $\tilde M$ have an eigenvalue which is positive but arbitrarily close to $0$, and then have some other diagonal matrix elements of $M$ (for example corresponding to $\expec[qppq]$) which are taken larger (tending to infinity as the smallest eigenvalue of $\tilde M$ tends to zero) keeping the matrix $M$ semi-definite.

Having given the limitations of SoS methods in this case, let us comment on one interesting calculation that does indicate how these methods could be used to reproduce perturbation theory without needing to consider the generalization of conjecture
\ref{conjrsos}.  We begin with some algebraic manipulations, but the final
result involves one heuristic step.  Let us explain:
consider
the operator
$$d\equiv Aq+ip+gBq^3,$$
where $A,B$ are real scalars.
We have
\be
\frac{d^\dagger d}{2}=\frac{A^2}{2}q^2 - \frac{3}{2}gBq^2
+ABgq^4
+g^2 B^2 \frac{q^6}{2}
-\frac{A}{2}.
\ee
Choosing
$$AB=1,$$
and
$$A^2-3gB=1,$$
we have
\be
\frac{d^\dagger d}{2}=H-\frac{A}{2}+g^2 B^2 \frac{q^6}{2},
\ee
or, equivalently,
\be
H=\frac{d^\dagger d}{2}+\frac{A}{2}-g^2 B^2 \frac{q^6}{2}.
\ee
Solving these equations for small $g$, we let $A=1+xg+yg^2+\cO(g^3),$
and
we need $A^2-3g/A=1,$ so (ignoring terms of order $\cO(g^3)$ on both sides) we have
$1+2x g +(x^2+2y)g^2-3g+3xg^2=1.$
Hence, $x=3/2$ and $y=-(x^2+3x)/2=-27/8$.
So,
$$
\frac{A}{2}=\frac{1}{2}+\frac{3g}{4}-\frac{27g^2}{16}+\cO(g^3).
$$
We also have that the expectation value of
$g^2 B^2 q^6/2$ \emph{in the ground state of the harmonic oscillator} (i.e., at $g=0$) is equal to $(15/16) g^2 B^2=(15/16)g^2$.
Since $27/16+15/16=42/16$,
we have
\be
H=
\frac{d^\dagger d}{2}+\frac{1}{2}+\frac{3g}{4}-\frac{42g^2}{16}-\Bigl( g^2 B^2 \frac{q^6}{2}-(15/16)g^2 B^2\Bigr)+\cO(g^3).
\ee

The expectation value of $g^2 B^2 \frac{q^6}{2}-(15/16)g^2 B^2$ vanishes in the ground state at $g=0$.
So, one should expect that, if the ground state at small with is close (within $\cO(g)$) of the ground state at $g=0$, then the expectation value of $g^2 B^2 \frac{q^6}{2}-(15/16)g^2 B^2$ in that ground state is $\cO(g^3)$.  Hence, under this assumption, we would have
that the ground state energy at nonzero $g$ is 
$\frac{1}{2}+\frac{3g}{4}-\frac{42g^2}{16}+\cO(g^3),$ which in fact agrees with
standard perturbation theory calculations.  Dealing with this in a precise way will require some extra steps, and we leave that for future work.  We expect that using conjecture \ref{conjrsos} it will be possible to prove bounds which agree with perturbation theory for any given order.

Finally, let us show that using the kind of generalization of conjecture \ref{conjrsos} it is possible to
at least reproduce first order perturbation theory.
Let
$$L=1+\delta q^4,$$
where the scalar $\delta$ will be chosen later.
We will write
$L^\dagger (H-\lambda_0) L$ as a sum-of-squares for $L=1/2+3g/4+\cO(g^2)$, certifying that the ground state energy is $\geq \lambda_0$.

Let us introduce notation: $A \geqsos B$ will mean that the operator $A-B$ is a sum of squares. So, our goal is to prove $L^\dagger (H-\lambda_0) L\geqsos 0$ for some $\lambda_0$ reproducing first order perturbation theory.

We have $q^4 H q^4\geqsos 0$.
So,
\be
L^\dagger (H-\lambda_0) L \geqsos
H-\lambda_0
+\delta \{p^2,q^4\}
+2\delta q^6
+2\delta g q^8-\delta^2 \lambda_0 q^8.
\ee
We have $\{p^2,q^4\}=pq^2p-12q^2$, so $\{p^2,q^4\}\geqsos -12q^2$.
Hence,
\be
L^\dagger (H-\lambda_0) L \geqsos
H-\lambda_0
-12\delta q^2
+2\delta q^6
+2\delta g q^8-\delta^2 \lambda_0 q^8.
\ee

Later, we will pick $\delta=\cO(g^2)$.
Then, for sufficiently small $q$, we have $2\delta g-\delta^2\geq 0$
so that under these assumptions
\be
L^\dagger (H-\lambda_0) L \geqsos
H-\lambda_0
-12\delta q^2
+2\delta q^6.
\ee
Note that $H-12\delta q^2=p^2/2+(1-24 \delta)q^2/2+gq^4,$
so $H-12\delta q^2$ is of the same form as $H$, except that there is a change in the coefficients.  Indeed, we have $H-12\delta q^2\geqsos (1-24 \delta) H$.  Remark: at this point tighter bounds are definitely possible.

So,
\be
L^\dagger (H-\lambda_0) L \geqsos
(1-24\delta) \Bigl( H-\frac{\lambda_0}{1-24\delta}
+\frac{2\delta q^6}{1-24\delta}\Bigr).
\ee

Above, we have shown that
\be
H=\frac{d^\dagger d}{2}+\frac{A}{2}-g^2 B^2 \frac{q^6}{2},
\ee
with scalars $A$ and $B$ given above.
So, choosing $\lambda_0/(1-24\delta)=A/2$ and $$4\delta q^6/(1-24\delta)=g^2 B^2,$$
we have shown that $L^\dagger (H-\lambda_0) L \geqsos 0$.
Note that this indeed gives $\delta=\cO(g^2)$ as needed.
So, this establishes the desired result for
$$\lambda_0=A/2(1+24\delta+\cO(\delta^2))= 1/2+3g/4+\cO(g^2).$$

\bibliographystyle{unsrturl}
\bibliography{refs}
\end{document}